
413 gives examples
\font\tenrm=cmr10
\font\ninerm=cmr9
\font\nineit=cmti9
\font\ninebf=cmbx9

\font\veinterm=cmr9 scaled 2000
\font\veintebf=cmbx9 scaled 2000

\font\eightrm=cmr8
\font\eightit=cmti8

\font\twelvebf=cmbx10 scaled 1200

\nopagenumbers
\hsize=6.67truein
\vsize=9truein

\veinterm

\def\r{{\bf r}}
\def\zeroz{\{z_i (\zeta)\} }
\def\zero{\{z_i\} }
\def\azeroz{\{A_i (\zeta)\} }

\baselineskip=12truept
\noindent{\eightit Journal of Superconductivity, Vol. 8, No. 1, 1995

\vskip .3truein
\noindent{\tenrm \bf
Presented at the Miami Workshop on High Temperature
Superconductivity, January 5-11, 1995}
\vskip .4truein

\noindent{\veintebf New Phases of HTS in High
Magnetic Field}

\vskip .2in
\twelvebf

\noindent{\bf Zlatko Te{\v s}anovi{\' c}${^1}$}
\vskip .1in
\eightrm

\quad\quad\quad\quad\quad\qquad\quad\quad{\nineit Received 15 January 1995}

\quad\quad\quad\quad\quad\qquad\quad\quad{\veintebf------------------------------------------------------}

\quad\quad\quad\quad\quad\qquad\quad\quad {\ninerm
Fluctuation behavior of HTS
in high magnetic field is studied within the Ginzburg-Landau}

\quad\quad\quad\quad\quad\qquad\quad\quad {\ninerm
theory. Landau level degeneracy of Cooper pairs
enhances fluctuations which destroy the}

\quad\quad\quad\quad\quad\qquad\quad\quad {\ninerm
familiar Abrikosov lattice. Instead,
a charge density-wave of Cooper pairs (SCDW) is}

\quad\quad\quad\quad\quad\qquad\quad\quad {\ninerm
the new low-temperature phase
of the theory. SCDW has no condensate, but
differs from}

\quad\quad\quad\quad\quad\qquad\quad\quad {\ninerm
the normal
state by a
periodic modulation of Cooper pair density. In presence of disorder,}

\quad\quad\quad\quad\quad\qquad\quad\quad {\ninerm
Abrikosov state is revived and both superconducting and
density-wave phases
are possible.}

\quad\quad\quad\quad\quad\qquad\quad\quad{\veintebf------------------------------------------------------}

\quad\quad\quad\quad\quad\qquad\quad\quad {\ninebf KEY WORDS:} {\eightrm
Type-II Superconductivity,
Charge Density Wave, Ginzburg-Landau Model.}

\vskip .35truein
\newdimen\fullhsize
\fullhsize=6.67truein \hsize=3.17truein
\def\fullline{\hbox to\fullhsize}

\let\lr=L \newbox\leftcolumn
\output={\if L\lr
   \global\setbox\leftcolumn=\columnbox \global\let\lr=R
  \else \doubleformat \global\let\lr=L\fi
  \ifnum\outputpenalty>-20000 \else\dosupereject\fi}
 \def\doubleformat{\shipout\vbox{\makeheadline
     \fullline{\box\leftcolumn\hfil\columnbox}
     \makefootline}}
  \def\columnbox{\leftline{\pagebody}}
\tenrm

\vskip.24truein
There has been much interest lately in the fluctuation behavior
of high temperature superconductors (HTS) and related systems.
Here I briefly review some of the recent progress on the
fluctuation problem in high magnetic fields. I also address the
relationship between the high- and the low-field regimes
of critical behavior.

Fluctuations of type-II superconductors
in magnetic field ${\bf H}$ are described
by the Ginzburg-Landau (GL) functional
$\int d^2rd^{D_p}\zeta F_{GL} [\psi (\r,\zeta)]$, where:
$$
F_{GL} = \sum_j \left[a_j (T,H)|\psi_j|^2 +
\gamma |{\bf \partial}_\zeta \psi_j|^2\right] +
{b\over 2}|\psi|^4~~.
\eqno(1)
$$
Here $a_j(T,H), b,$ and $\gamma$
are material-dependent parameters and $D_p$ is
the number of dimensions `along' ${\bf H}$.
An important feature of $F_{GL}$
is the formation of Landau levels (LLs) for Cooper pairs.
$\psi _j (\r,\zeta)$ is a component of the
fluctuating order parameter field belonging to
the $j$th LL.

The LL structure arises from the
quadratic part of $F_{GL}$ and is
independent on any particular representation of the
fluctuation problem: it is dictated by
gauge and spatial symmetries.
The quartic interaction term, however,
mixes different LLs and acts
to suppress the LL structure in
the fluctuation spectrum. It is convenient
to split the effect of the
quartic
\vskip.1truein
\noindent{\veintebf---------}

\noindent{\eightrm $^1$Department of Physics and Astronomy,
Johns Hopkins University, Baltimore, MD 21218, USA}

\topskip 4.125truein
\noindent
term into the intra-LL
and inter-LL correlations. At high fields,
$H\gg H_b$, where the
cyclotron gap between LLs is much larger
than the interaction term,
only the {\it intra}-LL
correlations are important and the LL structure
will be reflected in the theory. In this regime
keeping only the lowest LL suffices in capturing essential
features of the physics.
In the opposite limit of
low fields, $H\ll H_b$, the {\it inter}-LL
correlations become dominant and
the LL structure is suppressed
at long wavelengths. The crossover field separating
these two regimes is
$H_b\sim (\theta/16)(T/T_{c0})H_{c2}(0)$,
where $\theta$ is the Ginzburg fluctuation number,
$\theta\cong
2b H'_{c2}T_{c0}^2/
\phi_0a(0,0)^{3/2}\gamma^{1/2}$, where
$H'_{c2} = [dH_{c2}/dT]$ at $T=T_{c0}$.
This expression for $H_b$ was derived in
Ref. [1] by comparing the strength
of quartic correlations
in (1) with the cyclotron gap between LLs.
In HTS $\theta \sim 0.01-0.05$
and $H_b\sim 0.1 - 1$ Tesla.[1]

Here we are interested in high fields, $H\gg H_b$.
We keep only the LLL in $F_{GL}$ and assume
that $a_0, b,$ and $\gamma$ are
renormalized by fluctuations from higher LLs.
This defines the `renormalized' GL-LLL theory:
$$
F_{GL-LLL} = a_0 (T,H)|\psi|^2 +
\gamma |{\bf \partial}_\zeta \psi|^2 +
{b\over 2} |\psi|^4~~,
\eqno(2)
$$
with the constraint $\psi (\r,\zeta)\in$ LLL.
This GL-LLL theory exhibits dimensional reduction (DR) as a
consequence of the LL degeneracy: Within perturbation
expansion its properties appear related to the $D_p=D-2$-dimensional
GL theory in {\it zero} field [2]. Furthermore, the
pairing susceptibility exhibits
{\it exact} DR in the normal state, suggesting the absence of the
superconducting (Abrikosov) transition for $D<4$ [3].
Indeed, the high order perturbation
expansion for $D=2,3$ shows no indication
of transition to the Abrikosov vortex lattice [2].
Similarly, the low-temperature
\topskip 0truein
\noindent
harmonic expansion starting from the perfect lattice
is plagued by infrared divergences for $D=2,3$ ($D_p =0,1$) [4].
As stressed by Moore, these results must be interpreted
as terminal instability of the Abrikosov
vortex lattice to fluctuations [4]. Since small fluctuations
destabilize the low-temperature state predicted by the mean-field
theory, it was ``natural" for Moore to conclude that GL-LLL theory
exhibits no phase transitions for
$D=2,3$ and describes a single, normal
phase at all temperatures [4].
In contrast, for $D>4$ ($D_p >2$)
the Abrikosov transition is restored and is likely
first order [5].

The complete picture is more complex.
It has been proposed that the GL-LLL theory
has actually {\it two different} mechanisms which can lead to
a phase transition at low-temperatures [3].
The first mechanism is just the familiar one discussed
by Moore: it produces the Abrikosov's  vortex lattice state,
renormalized by fluctuations.
Such state has off-diagonal
long range order (ODLRO) or quasi ODLRO in the one-body
density-matrix and is characterized
by a diverging superconducting pairing
susceptibility, $\chi_{sc}$. The lower critical dimension
for this transition is indeed four ($D_p =2 $).
There is, however,
another mechanism present in the theory. This new mechanism
is exclusively due to the
{\it intra}-LL {\it amplitude} correlations and
is therefore operational only at high fields ($H\gg H_b$).
It leads to formation of the charge density-wave
of Cooper pairs in the plane perpendicular to ${\bf H}$.
The lower critical dimension in this case is two and
it is thus the density-wave mechanism which produces
phase transitions in physical systems, at least
in the weak pinning regime. This new phase transition has been
observed in numerical simulations
[6] and appears weakly discontinuous [6,7].
Consequently, the ``natural" conclusion is not:
the GL-LLL theory does have a phase transition both in
2D and 3D. It is, however, a new charge density-wave transition,
unrelated to the Abrikosov vortex lattice. The charge
density-wave state is `normal' in the sense that it
has no ODLRO -- Superconducting correlations are short
ranged. For $D>4$ ($D_p >2$) both density-wave and
Abrikosov vortex lattice transitions are possible.

The differences between the two mechanisms and
between the Abrikosov lattice and the charge
density-wave are subtle
but important.  The Abrikosov vortex lattice
is produced by the BCS-type
mechanism which, in this context,
we can think of as the Bose condensation [BC] of Cooper
pair field $\psi$ into a single quantum state $\Psi_A\in$ LLL.
Since all states in the LLL can be
represented [in symmetric gauge] as
$\psi (\r,\zeta)=\phi (\zeta )\prod_i[z-z_i(\zeta)]$
the condensate wave function can be expressed as
$\Psi_A \equiv \langle\psi (\r,\zeta)\rangle
=\Phi (\zeta ) \prod_i[z-A_i(\zeta)]$, where $\Phi (\zeta) =
\langle\phi (\zeta)\rangle$ and $\azeroz$
are the {\it Abrikosov vortices}.
What matters here
is that the {\it effective} $a_0$, which includes the
renormalization from the quartic interaction, goes to
zero for some of the states in the LLL.
So, this mechanism is tied to the renormalized
quadratic term in the GL-LLL theory. Note that
BC and ensuing ODLRO trivially imply
breaking of spatial symmetries since
all $\psi (\r,\zeta)\in$ LLL are non-uniform.
The second mechanism arises from the {\it quartic}
term and is a pure correlation effect. These
amplitude correlations are a direct consequence of the
LLL constraint. Their effect is measured
by the ratio $\int {d^2r\over \Omega}|\psi |^4/
[\int {d^2r\over \Omega}|\psi |^2]^2\equiv
\beta_A (\zeroz)$.
The quartic term would ideally like to
force $\beta_A (\zeroz )$ to unity but that is possible
only for the uniform amplitude which is
not available in the LLL. The result are
strong lateral amplitude correlations of $\zeroz$--When
these positional correlations force $\langle\beta_A\rangle$
below certain critical value ($\sim 1.2$ in 2D)
the average pair density
$\langle|\psi (\r,\zeta)|^2\rangle$ cannot remain
uniform and develops a weak modulation
in the xy-plane. Thus, the second mechanism leads to
formation of the charge density-wave of Cooper
pairs (SCDW). The SCDW transition generally does not
induce BC and superconducting ODLRO. The SCDW is
still completely incoherent, like the normal state, and
$\Phi (\zeta) = 0$!

While in general both BC and SCDW mechanisms are at work
in the GL-LLL theory,
they are not related in any simple fashion. For example,
one can have a situation in which a perfect Abrikosov vortex lattice
is formed below some BC temperature in absence
of any positional correlations among $\zeroz$.
In this case, the
normal state above this BC temperature and the
Abrikosov votex lattice below are both
uncorrelated {\it liquids} in terms of $\{z_i(\zeta)\}$.
Conversely, we can have a perfect, highly-correlated,
defect-free triangular solid of $\{z_i(\zeta)\}$
with a {\it positive} effective $a_0$ and
consequently no BC and only a short-range superconducting order.
Furthermore, having some information about
positional correlations
of $\{z_i(\zeta)\}$ does not easily translate into
any particular information about superconducting correlations.
This can be traced back to
the fundamental difference between $\zeroz$ and $\azeroz$:
the former are {\it microscopic} variables like creation
and annihilation operators and are fluctuating strongly,
the latter are {\it macroscopic} objects which
either remain fixed in the thermodynamic equilibrium
or change slowly in response to external
fields and currents. Fluctuations of $\zeroz$ determine
$\Phi (\zeta)$ but only in a highly
convoluted way. Similarly, the
relationship between Abrikosov vortices
$\azeroz$ and $\zeroz$ is anything but simple.
In general, to relate the latter to the former one
needs the knowledge of the effective
$a_0$ plus {\it all} the density
correlations of $\{z_i(\zeta)\}$, clearly
a formidable problem.

There are some
exceptions from this rather hopeless situation,
the simplest being the low temperature limit.
In that case individual
$\zeroz$ can be viewed as `bound' to individual
$\azeroz$ and executing only small oscillations around
them. We are then justified in calculating
the superconducting correlator within the harmonic
approximation for the motion of $\zeroz$.
As the temperature increases, however, $\zeroz$
`unbind' from $\azeroz$ and we are back to the above
difficult problem. It is precisely this `unbinding'
that characterizes the strong fluctuation regime
of the GL-LLL theory and it is precisely in this
strong fluctuation regime that various phase
transitions of interest take place.
Just how the BC and SCDW mechanisms interact in the
strong fluctuation regime and for various
physical situations is
the key problem in this field.

There has been recent progress in understanding at
least some aspects of this difficult problem.
In 2D and in highly anisotropic layered systems
one is far below $D=4$ and the superconducting
correlations are effectively zero-dimensional:
For example, in a 2D film the exact
superconducting correlator is $\langle\psi (\r)
\psi^* (0)\rangle \propto \langle |\psi |^2\rangle
\exp ( - r^2/4\ell ^2 )$, for all temperatures.
In this case we can concentrate on the SCDW
mechanism and forget about BC. Above $H_{c2}(T)$
the quartic correlations are weak and $\langle\beta_A\rangle
\sim 2$. As we move below $H_{c2}(T)$, the
lateral correlations increase, gradually forcing
$\langle\beta_A\rangle$ toward 1.159, its minimum
value within the LLL. Let us consider different configurations
of $\zero$ which all give some particular value
for $\beta_A\{ z_i\}$ (uniform translations
and rotations excluded). Only such configurations
contribute to the thermodynamic sum when
$\beta_A\{ z_i\}=\langle\beta_A\rangle$. For
$\langle\beta_A\rangle > \beta_{Ac}$ the average
of $|\psi (\r)|^2$ over all configurations of
$\zero$ remains uniform. However, for
$\langle\beta_A\rangle < \beta_{Ac}$,
$\langle|\psi (\r)|^2\rangle$ is weakly modulated
with the period $\sim \ell\equiv\sqrt{c/2eH}$ but
otherwise has no simple relation to the periodicity
of the Abrikosov vortex lattice. In 2D $\beta_{Ac}\sim 1.2$.
This 2D SCDW state
has no superconducting phase coherence at all
and is perfectly `normal' in the sense of not
having a condensate. The mean-field description
of SCDW has been constructed in Ref. [8] using
density-functional [DF] theory. Such mean-field description
is, of course, completely different
from that of Abrikosov since, from the standpoint
of Abrikosov solution, SCDW is a pure
fluctuation-induced phase. While the DF theory of the normal-SCDW
transition is quite simple, it produces a very
good quantitative agreement with numerical simulations [6,7].

What is the phenomenology of the SCDW and how can we
distinguish this new phase from the normal metal
and the Abrikosov vortex lattice state? First, the SCDW
is not a superconducting phase, at least in
quasi 2D systems, like highly anisotropic layered
materials. So, the resistivity is finite, there is
no macroscopic phase coherence, no Josephson effect,
and no flux quantization. There is, however, a
periodic lateral modulation of the Cooper pair
density with a period which is in general {\it different}
from that of the Abrikosov vortex lattice.
A key property of the
high-field limit is that this periodicity
can be used to reconstruct the full
superconducting correlator: If a weakly-modulated
$\langle|\psi (\r)|^2\rangle$ is known,
analyticity properties of the LLL
allow one to determine $\langle\psi (\r)\psi ^*(\r')\rangle$.
This property is {\it unique} to the high-field limit.
The SCDW will lead to
the modulation of pseudo-gap in the electronic
density of states which can be observed in an STM
tunneling experiment. A two-tip STM experiment, with
two tips being able to move in- and out-of-phase,
could simultaneously measure both the SCDW density
modulation and range of superconducting phase correlations,
the latter giving rise to the Josephson current
between the tips. This would be a direct experimental
test of analytical predictions of the GL-LLL theory.

The above discussion illustrates SCDW
mechanism at work, producing a novel
fluctuation-induced state in type-II superconductors.
Alternatively, we could look
for the situation where the BC
mechanism dominates. An academic playground for
this is provided by the $D>4$-dimensional GL-LLL
theory where both Abrikosov lattice and
SCDW transitions are possible. In this case, we
can study models where only the BC mechanism is
operational, like the $1/N$-expansion of the
vector GL-LLL theory, in the $N\to\infty$ limit.
The real world (un)fortunately is $D<4$-dimensional
and only the SCDW mechanism survives.
However, we can still beat dimensional reduction (DR)
if the translational symmetry in the xy-plane is broken
not spontaneously, like in SCDW, but explicitly, by some
external potential, $V(\r)$.  The LL degeneracy is lifted
by such external potential and DR
is now preempted at temperatures less than $V(\r)$.
It is now possible for the BC mechanism to produce
a finite temperature superconducting transition even
in the {\it absence} of SCDW correlations.
There will be two sources of such LL broadening in
real superconductors: disorder arising from impurities
and pinning centers and periodic potential of the
underlying crystalline lattice. For situation of interest
disorder is usually more important. I consider here
the so-called columnar disorder, which is both theoretically
pleasing and of practical importance. The GL functional
(2) now becomes:
$$\left[a_0+
\lambda \sum_{i} \delta (\vec{r}-\vec{r_{i}})\right]|\psi|^{2}
+ \gamma |\partial_{\zeta}\psi|^2 + {b\over 2}
|\psi|^{4}~~,
\eqno(3)
$$
where $\lambda>0$ characterizes local
suppression of superconductivity at
defect sites.  The magnetic field is
assumed to be parallel to columnar defects,
the effective potentials of all
defects the same and well-represented by delta-functions.
Random variables in the problem are 2D coordinates of
defects, $\{\r_{i}\}$. We assume that columns of
damaged superconductor are distributed according to
the Poisson law
$P_{N}(\r_{1},...\r_{N})=
(e^{-\rho \Omega} \rho^{N})/N! $
where $P_{N}$ is the probability
for finding $N$ impurities at the
positions $\r_{1},...\r_{N}$,
$\Omega$ is the lateral area of the system
and $\rho$ is the concentration of defects.

Since here we are interested in the BC mechanism
we can ignore lateral amplitude
correlations and include their
effect only in an average way, by replacing
$\int {d^2r\over \Omega}|\psi |^4\to\langle\beta_A\rangle
[\int {d^2r\over \Omega}|\psi |^2]^2$, where
$\langle\beta_A\rangle$ is the thermodynamic
average, directly in $F_{GL}$. This eliminates
the SCDW mechanism and the resulting model
can be solved {\it exactly} [9]. This is a non-trivial
point: The exact solution is possible because the
analyticity of the LLL states allows one to
find the density of states in presence of
disorder. This is all one needs to compute the
exact thermodynamics of the model.
Similarly, the ``glassy" superconducting correlator, defined as
$|\langle\psi (\r,\zeta)\psi^* (0,0)\rangle|^2$
averaged over disorder, can be calculated
at long distances by appealing to anomalous diffusion
of LLL states. Since we neglected SCDW correlations,
the critical exponents of this superconducting transition
are `classical' but the solution
exhibits interesting `dimensional transmutation':
The behavior of specific heat, order parameter and
correlation length in this problem is related to the
O(2$N$) vector model in the limit $N\rightarrow \infty$
and in the {\it effective} dimension $D_{\rm eff}=2f-1$.
Here $f\equiv\rho 2\pi\ell^{2}$ measures density
of defects relative to density of zeros.
The continuous change of $D_{\rm eff}$ with $f$
reflects the extent to which disorder has lifted
the LLL degeneracy.

It should be stressed that the above glassy superconducting
transition has been {\it induced} by disorder. Typically,
we study the effect of disorder on transitions in a clean system
and that effect is often detrimental. Here, however,
the LL degeneracy and dimensional reduction (DR) {\it prevent}
the superconducting transition in clean systems.
Disorder broadens LLL and relieves the frustration
that inhibited BC for $D<4$.
We might be tempted to think of this transition as simply pinning
the vortices down so they don't move around, but we
should resist the temptation! $\zeroz$ are still rapidly moving all
over the place as evidenced by $\langle\beta_A\rangle\sim 2$,
both above and below the transition. The zeroes
simply spend little more time on the average
in certain regions than in others, as dictated
by disorder.  This is sufficient to favor large occupancy of
certain states in the LLL and induce BC.
While ignoring SCDW correlations
was obvious oversimplification, it is still useful to
have an exact solution [9] illustrating how the BC
mechanism and superconducting state are
revived when DR is suppressed by LLL
broadening. The condensate wave function
$\Psi_A \equiv \langle\psi (\r,\zeta)\rangle
=\Phi (\zeta ) \prod_i[z-A_i(\zeta)]$ has Abrikosov vortices
$\azeroz$ distributed in some random fashion,
dictated by disorder.

Finally, we can put both
the SCDW and superconductivity (BC) together.
At present, there is no framework within which
problems where both mechanisms are operational
can be solved. An interesting possibility here is that
the anomalous nature of diffusive LLL density correlations
might lead to novel forms of critical behavior.

The high-field fluctuation behavior is characterized by a
LL degeneracy and its splitting by {\it intra}-LL correlations
and/or external potential.
As the field is reduced there is a qualitative change in this
picture. The cyclotron gap shrinks and the LL mixing produced
by the quartic interaction becomes very strong.
These {\it inter}-LL correlations are the key to correctly
describing the low-field, $H\ll H_b$, behavior.
Physically, the low energy fluctuation spectrum undergoes
a qualitative change: While in the high-field regime the
only relevant degrees of freedom were {\it field-induced}
vortices $\zeroz$, in the low-field limit there are
fluctuations of {\it thermally-induced}
vortex-antivortex pairs, vortex loops, etc. These
{\it zero vorticity} fluctuation modes originate
from high LLs, their `mass' being reduced by strong
inter-LL mixing. An important question here is
whether these zero vorticity modes condense at
some finite (low) field. This would imply the existence
of a line of finite field phase transitions terminating
in the zero field critical point at $T=T_{c0}$.
It was recently shown that such finite field transition
indeed occurs in certain continuum models [10].
This novel ordering is not associated with field-induced vortices,
which remain in the fluid state both above and below the
transition. Rather, it is a subtle form of off-diagonal
ordering of the zero vorticity degrees of freedom.
Such ordering signifies the qualitative difference
between the high- and the low-field regimes of
critical behavior.

I would like to thank my collaborators I. F. Herbut,
R. {\v S}{\' a}{\v s}ik, and D. Stroud for many
useful and productive discussions. The work reported
here was supported in part by the NSF grant
DMR-9415549.

\vskip.2in
\noindent{\bf{REFERENCES}}
\vskip.2in

\item{1. }
Z. Te{\v s}anovi{\' c} and A. V. Andreev, Phys. Rev. B {\bf 49},
4064 (1994).

\item{2. }
G. J. Ruggeri and D. J. Thouless, J. Phys. {\bf F} {\bf 6}, 2063 (1976);
E. Br{\' e}zin, A. Fujita, and S. Hikami, Phys. Rev. Lett.
{\bf 65}, 1949 (1990).

\item{3. }
Z. Te\v{s}anovi\'{c}, Physica (Amsterdam) C {\bf 220}, 303 (1994);
Z. Te\v{s}anovi\'{c}, Phys. Rev. B {\bf 44}, 12635 (1991).

\item{4. }
M. A. Moore, Phys. Rev. B {\bf 45}, 7336 (1992);
K. Maki and H. Takayama, Prog. Theor. Phys. {\bf 46},
1651 (1971).

\item{5. }
E. Br{\' e}zin, D. R. Nelson, and A. Thiaville, Phys. Rev. B
{\bf 31}, 7124 (1985).

\item{6. }
Z. Te\v{s}anovi\'{c} and L. Xing, Phys. Rev. Lett. {\bf 67}, 2729 (1991).

\item{7. }
Y. Kato and N. Nagaosa, Phys. Rev. B {\bf 47}, 2932 (1993);
J. Hu and A. H. MacDonald, Phys. Rev. Lett. {\bf 71}, 432 (1993);
R. {\v S}{\' a}{\v s}ik and D. Stroud, Phys. Rev. B {\bf 49},
16074 (1994).

\item{8. }
I. F. Herbut and Z. Te{\v s}anovi{\' c}, Phys. Rev. Lett.
{\bf 73}, 484 (1994).

\item{9. }
Z. Te{\v s}anovi{\' c} and I. F. Herbut, Phys. Rev. B
{\bf 50}, 10389 (1994).

\item{10. }
Z. Te{\v s}anovi{\' c}, preprint (1994).

\end